\documentclass[dvips]{article}
\newcommand{\doublespacing}{\let\CS=\@currsize\renewcommand{\baselinesstrech}
{2.0}\tiny\CS}

\begin{document}

\textwidth 16cm
\newcommand{\bd}{\begin{document}}
\newcommand{\ed}{\end{document}}
\newcommand{\bc}{\begin{center}}
\newcommand{\ec}{\end{center}}
\newcommand{\bfr}{\begin{flushright}}
\newcommand{\efr}{\end{flushright}}
\newcommand{\lt}{\left}
\newcommand{\rt}{\right}
\newcommand{\vs}{\vspace}
\newcommand{\hs}{\hspace}
\newcommand{\beq}{\begin{equation}}
\newcommand{\eeq}{\end{equation}}
\newcommand{\lb}{\linebreak}
\newcommand{\pb}{\pagebreak}
\newcommand{\mb}{\makebox}
\newcommand{\fb}{\framebox}
\newcommand{\mc}{\multicolumn}
\newcommand{\ben}{\begin{enumerate}}
\newcommand{\een}{\end{enumerate}}
\newcommand{\bit}{\begin{itemize}}
\newcommand{\eit}{\end{itemize}}
\newcommand{\ol}{\overline}
\newcommand{\un}{\underline}
\newcommand{\lefq}{\lefteqn}
\newcommand{\ba}{\begin{array}}
\newcommand{\ea}{\end{array}}
\newcommand{\beqa}{\begin{eqnarray}}
\newcommand{\eeqa}{\end{eqnarray}}
\newcommand{\beqas}{\begin{eqnarray*}}
\newcommand{\eeqas}{\end{eqnarray*}}
\newcommand{\bfg}{\begin{figure}}
\newcommand{\efg}{\end{figure}}
\newcommand{\bds}{\begin{displaymath}}
\newcommand{\eds}{\end{displaymath}}
\newcommand{\btb}{\begin{tabbing}}
\newcommand{\etb}{\end{tabbing}}
\newcommand{\para}{\parallel}
\newcommand{\pad}{\partial}
\newcommand{\nn}{\nonumber}
\newcommand{\la}{\leftarrow}
\newcommand{\ra}{\rightarrow}
\newcommand{\lgla}{\longleftarrow}
\newcommand{\lgra}{\longrightarrow}
\newcommand{\La}{\Leftarrow}\newcommand{\Ra}{\Rightarrow}
\newcommand{\Lra}{\Leftrightarrow}
\newcommand{\Lgla}{\Longleftarrow}
\newcommand{\Lgra}{\Longrightarrow}
\newcommand{\bm}{\boldmath}
\newcommand{\lan}{\langle}
\newcommand{\ran}{\rangle}
\renewcommand{\a}{\alpha}
\renewcommand{\b}{\beta}
\newcommand{\g}{\gamma}
\newcommand{\G}{\Gamma}
\renewcommand{\d}{\delta}
\newcommand{\eps}{\epsilon}
\newcommand{\Th}{\Theta}
\newcommand{\s}{\sigma}
\newcommand{\lam}{\lambda}
\newcommand{\D}{\Delta}
\newcommand{\vare}{\varepsilon}
\newcommand{\pr}{\prime}
\newcommand{\ro}{\rho}
\newcommand{\nab}{\nabla}
\newcommand{\m}{\mu}
\newcommand{\n}{\nu}
\newcommand{\Sg}{\Sigma}
\newcommand{\p}{\pi}
\newcommand{\R}{I\!\!R}
\newcommand{\om}{\omega}
\newcommand{\Om}{\Omega}
\newcommand{\ze}{\zeta}
\newcommand{\vart}{\vartheta}
\newcommand{\tri}{\triangle}
\newcommand{\f}{\frac}
\newcommand{\iny}{\infty}
\newcommand{\pro}{\propto}
\bc {\huge {\bf Higher order intertwining approach to quasinormal modes}} \ec

\vs{1cm}

\bc
{\it T. Jana{\footnote {e-mail : tapas$_{-}$r@isical.ac.in} and P. Roy{\footnote{e-mail : pinaki@isical.ac.in}}\\
Physics \& Applied Mathematics Unit \\
Indian Statistical Institute \\
Kolkata - 700 108, India.}} \ec
\vs{4.5cm}

\bc {\large {\un{Abstract}}} \ec Using higher order intertwining
operators we obtain new exactly solvable potentials admitting
quasinormal mode (QNM) solutions of the Klein-Gordon equation. It is
also shown that different potentials
exhibiting QNM's can be related through nonlinear supersymmetry.  \\
\pb

\section {Introduction}
Quasinormal modes (QNM) are basically discrete complex freequency
solutions of real potentials. They appear in the study of black
holes and in recent years they have been widely studied
\cite{kokko}. Interestingly QNM's have also been found in
nonrelativistic systems \cite{ho}. However as in the case of bound
states or normal modes (NM) there are not many exactly solvable
potentials admitting QNM solutions and often QNM frequencies have to
be determined numerically or other approximating techniques such as
the WKB method, phase integral method etc. Consequently it is of
interest to obtain new exactly solvable potentials admitting such
solutions.

In the case of NM's or scattering problems a number of methods based on intertwining
technique e.g, Darboux algorithm \cite{dar}, supersymmetric quantum
mechanics (SUSYQM) \cite{khare} etc. have been used successfully to
construct new solvable potentials. Usually the intertwining
operators are constructed using first order differential operators.
However in recent years intertwining operators have been generalised
to higher orders \cite{andrianov,plyush,fernan,samsonov,aoyama} and
this has opened up new possibilities to construct a whole new class
of potentials having nonlinear symmetry. In particular use of higher
order intertwining operator or higher order Darboux algorithm leads
to nonlinear supersymmetry.

QNM's are associated with outgoing wave like behaviour at spatial
infinity and unlike normal modes (NM's), the QNM wave functions have
rather unusual characteristic (for example, wave functions diverging
at both or one infinity) \cite{leung}. Such open systems have been
studied using (first order) intertwining technique \cite{anderson}.
Recently it has also been shown that open systems can be described
within the framework of first order SUSY \cite{leung}. Here our
objective is to examine whether or not intertwining method based on
higher order differential operators can be applied to open systems.
For the sake of simplicity we shall confine ourselves to second
order intertwining operators (second order Darboux formalism) and it
will be shown that the second order Darboux algorithm can indeed be
applied to models admitting QNM's although not exactly in the same
way as in the case of NM's. In particular we shall use the second
order intertwining operator to the P\"{o}schl-Teller potential to
construct several new potentials admitting QNM solutions. It will
also be shown that such potentials may be related to the
P\"oschl-Teller potential by second order SUSY. The organisation of
the paper is as follows: in section 2, we present construction of
new potentials using second order intertwining operators; in section
3, nonlinear SUSY underlying the potentials is shown and finally
section 4 is devoted to a conclusion.

\section{Second order intertwining approach to quasinormal modes}
Two Hamiltonians $H_0$ and $H_1$ is said to be intertwined by an
operator $L$ if \beq LH_0 = H_1L \eeq Clearly if $\psi$ is an
eigenfunction $H_0$ with eigenvalue $E$ then $L\psi$ is an
eigenfunction of $H_1$ with the same eigenvalue provided $L\psi$
satisfies required boundary conditions. If $L$ is constructed using
first order differential operators then intertwining method is
equivalent to Darboux formalism or SUSYQM. In particular if $V_0$ is
the starting potential and $L = {\displaystyle \f{d}{dx}+W(x)}$,
then the isospectral potential is $V_1 = V_0 + 2{\displaystyle
\f{dW}{dx}}$ \cite{khare}. Similarly if $L$ is generalised to higher
orders then it is equivalent to higher order Darboux algorithm or
higher order SUSY. \vspace{.15cm}

Let us now consider $L$ to be a second order differential operator of the form \cite{fernan}
\beq
\ba{lcl}
L &=& {\displaystyle \f{d^2}{dx^2} + \beta(x)\f{d}{dx} + \gamma(\beta)}\\\\
\beta(x) &=& {\displaystyle -\f{d}{dx}~log~ W_{i,j}(x)}\\\\
\gamma(\beta) &=& {\displaystyle -\f{\beta''}{2\beta} + \left(\f{\beta'}{2\beta}\right)^2 +
\f{\beta'}{2} + \f{\beta^2}{4} - \left(\f{\omega_i^2-\omega_j^2}{2\beta}\right)^2}\\
\ea \label{L} \eeq where $\psi_i$ and $\psi_j$ are eigenfunctions of
$H_0$ corresponding to the eigenvalues $\omega_i^2$ and $\omega_j^2$
and $W_{i,j} = (\psi_i\psi_j'-\psi_i'\psi_j)$ is the corresponding
Wronskian. Then the isospectral partner potential $V_2(x)$ obtained
via second order Darboux formalsim is given by \beq V_2(x) = V_0(x) -
2\f{d^2}{dx^2}\log W_{i,j}(x)\label{v2} \eeq The wave functions
$\psi_i(x)$ and $\phi_i(x)$ corresponding to $V_0(x)$ and $V_2(x)$
are connected by \beq \phi _k (x) = \displaystyle{L \psi _k (x) = \f{1}{W_{i,j}(x)}}\left|\ba{ccc} \psi_i & \psi_j & \psi_k\\ \psi'_i & \psi'_j & \psi'_k\\ \psi''_i & \psi''_j & \psi''_k\\ \ea\right|,~~ i,j\neq k\label{phi} \eeq  The eigenfunctions
obtained from $\psi_{i}$ and $\psi_{j}$ are given by \beq f(x)\propto
\f{\psi_{i}(x)}{W_{i,j}(x)}~~,~~g(x)\propto
\f{\psi_{j}(x)}{W_{i,j}(x)} \label{phiij} \eeq It may be noted
that in the case of normal modes, the new potential would be free of
any new singularities if the Wronskian $W_{i,j}(x)$ is nodeless.
This in turn requires that the Wronskian be constructed with the
help of consecutive eigenfunctions (i.e, $j=i+1$). Also, the
eigenfunctions $f(x),g(x)$ in (\ref{phiij}) are not acceptable
because they do not satisfy the boundary conditions for the normal
modes and in any case they are not SUSY partners of the
corresponding states in the original potential. Thus in the case of
normal modes the spectrum of the new potential is exactly the same
as the starting potential except for the levels used in the
construction of the Wronskian. However we shall find later that not
all of these results always hold in the case of QNM's.
\vspace{.25cm}

Let us now consider one dimensional Klein-Gordon equation of the
form \cite{leung} \beq [\partial_t^2-\partial_x^2 + V(x)]\psi(x,t) =
0\eeq\label{kg1} The corresponding eigenvalue equation reads \beq
H\psi_n = \omega_n^2 \psi_n(x),~~~~H = -\f{d^2}{dx^2} +
V(x)\label{qnm} \eeq  The QNM solutions of the equation (\ref{qnm})
are characterised by the fact that they are either (1) increasing at
both ends (II)  (2) increasing at one end and decreasing at the
other (ID,DI). The wave functions decreasing at both ends (DD)
correspond to bound states or NM's. In the case of QNM's the
eigenvalues ($\omega_n^2$) may be complex or real and negative. If
$Re( \omega_n)\neq 0$ then the SUSY formalsim can not be applied
since in that case the superpotential $W(x)$ becomes complex and
consequently one of the partner potential becomes complex. So we
shall confine ourselves to the case when $Re(\omega_n)=0$ i.e,
$\omega_n^2$ are real and negative. We would also like to mention
that in case Eq.(\ref{qnm}) is to be interpreted as a Schr\"odinger
equation one just has to consider the replacement
$\omega_n^2\rightarrow E_n$.

There are a number of potentials which exhibit QNM's. A potential in
this category is the inverted P\"oschl-Teller potential. This
potential is used as a good approximation in the study of
Schwarzschild black hole and it is given by \beq V_0(x) = \nu
sech^2x\label{qnm1} \eeq Eq.(\ref{qnm}) for the potential
(\ref{qnm1}) can be solved in different ways. One of the simplest
way is to apply the shape invariance criteria \cite{khare} and the
solutions are found to be \cite{mashhoon,blome,ferrari} \beq
\omega_n^\pm = -i(n - A^\pm),~ A^\pm = -\f{1}{2} \pm
\sqrt{\f{1}{4}-\nu} = -\f{1}{2} \pm q\label{omegan} \eeq \beq
\psi_n^\pm(x) =
(sechx)^{(A^\pm-n)}{_2F_1}(\f{1}{2}+q-i\omega_n^\pm,\f{1}{2}-q-i\omega_n^\pm,1-i\omega_n^\pm,\f{1+tanhx}{2})\label{qnm2}
\eeq We note that the behaviour of the wave functions (\ref{qnm2})
i.e, whether they represent a NM or QNM depends on the value of the
parameter $\nu$. For $\nu\in (0,1/4)$ i.e,
$A^+\in(-1/2,0),A^-\in(-1,-1/2) $, the wave functions represent
outgoing waves and are of the type (II). For $\nu<0$ (i.e, $A^+>0,
A^-<0$) then the wave functions represent NM's when $n<A^+$ while
for $n>A^+$ they are QNM's. On the other hand the wave functions are
always QNM's corresponding to $\omega_n^-$. It may be noted that for
the QNM's the wave functions (\ref{qnm2}) for even $n$ are nodeless
while those for odd $n$ have exactly one node at the orign. This
behaviour of the wave functions is quite different from those
occuring in the case of NM's. Using the procedure mentioned above we
shall now construct new exactly solvable potentials admitting QNM
solutions.

\subsection{Construction of isospectral partner potential using NM's}

{\underline{\bf Case 1. $V_0$ with two NM's:}}~ In order to apply
the second order intertwining approach one may start with a
potential $V_0(x)$ admitting (1) at least two NM's and the rest
QNM's or (2) only QNM's. We begin with the first possibility. Thus
we consider $\nu=-5.04$ so that there are two NM's. In this case we
obtain from (\ref{omegan}) $A^\pm=1.8,-2.8$. Thus the NM's
correspond to $\omega_0^+=1.8i,\omega_1^+=0.8i$ and are given by
\beq \psi_0^+(x) = (sech~x)^{A^+}~~,~~\psi_1^+(x) =
(sech~x)^{(A^+-1)}{_2F_1}(2A^+,-1,A^+,\f{1+tanhx}{2}) \eeq The QNM's
in this case correspond to the frequencies $\omega_n^+, n=2,3,...$
and are given by $\psi_n^+(x)$. Also for $\omega_n^-, n=0,1,2,...$
there is another set of QNM's and the corresponding wave functions
are given by $\psi_n^-(x)$. In this case $A^-<0$ and consequently
there is no NM.

\vspace{.25cm}

Let us construct a potential isospectral to (\ref{qnm1}) using the
NM frequencies $\omega_0^+$ and $\omega_1^+$. Then from (\ref{qnm2})
the Wronskian $W_{0,1}^+$ is found to be \beq W_{0,1}^+ = -(sech
x)^{2A^+-1}\label{w01} \eeq Clearly $W_{0,1}^+$ does not have a
zero. In this case the new potential $V_2^+(x)$ is free of
singularities and is given by \beq V_2^+(x) = V_0(x) -
2\f{d^2}{dx^2}\log W_{0,1}^+(x) = -(1-A^+)(2-A^+)sech^2x\label{v01}
\eeq Using the value of $A^+$ it is easy to see that the new
potential $V_2^+(x)$ in (\ref{v01}) does not support any bound state
but only QNM's. This is also reflected by the explicit expressions
for the wave functions. Using (\ref{phiij}) we find \beq f^+(x) =
(sech x)^{(1-A^+)}~~,~~g^+(x)= -(sech x)^{-A^+} tanhx\label{phi01}
\eeq From (\ref{phi01}) it follows that the above wave functions are
QNM's corresponding to $-\omega_1^+$ and $-\omega_0^+$ respectively.
Note that these two QNM's are new and were not present in the
original potential. This in fact is where the behaviour of the new
potential is different from the usual case. In the case of
potentials supporting only NM's the wave functions $f^+(x), g^+(x)$
obtained through (\ref{phiij}) do not have acceptable behaviour.
However in the present case both these wave functions become QNM's
instead of NM's and they have acceptable behaviour at $\pm\infty$ as
can be seen from (\ref{phi01}) as well as from figure 1. The other
wave functions $\phi_n^+(x), n=2,3,...$ corresponding to QNM
frequencies $\omega_n^+=-i(n-A^+)$ can be obtained using (\ref{phi})
and are given by \beq \ba{l}
\phi_n^+(x)= (sech x)^{(A^+-n)}\left[(n-1)n~F_n~tanh^2x\right.\\\\
\left.+ c_1(2n-3)~F_{n+1}~sech^2x~tanhx \right.\\\\+  \left.c_1c_2~F_{n+2}~sech^4x\right],~~
n=2,3,...\label{phin}
\ea
\eeq
where
\beq
\ba{lcl}
c_1 &=&\displaystyle{ -\f{n(2A^+-n+1)}{2(A^+-n+1)}~~,~~c_2~=~   \f{(-n+1)(2A^+-n+2)}{2(A^+-n+2)}}\\\\
F_n &=& {_2 F_1}(-n,2A^+-n+1,A^+-n+1,\f{1+tanhx}{2})\\\\
F_{n+1} &=& {_2 F_1}(-n+1,2A^+-n+2,A^+-n+2,\f{1+tanhx}{2})\\\\
F_{n+2} &=& {_2F_1}(-n+2,2A^+-n+3,A^+-n+3,\f{1+tanhx}{2})\\
\ea
\eeq

To see the nature of the wave functions (\ref{phin}) we have plotted
$\phi_2^+(x)$ and $\phi_3^+(x)$ in figure 1. From the figure it can
be seen that these wave functions are indeed QNM's and for even $n$
they do not have nodes while for odd $n$ they have one node at the
origin. We would like to point out that the new potential $V_2^+(x)$
has two more QNM's than $V_0(x)$. Thus except for two additional
QNM's, the QNM frequencies $\omega_n^+$ is common to both $V_0(x)$ and
$V_2^+(x)$. We now examine the second set of solutions corresponding
to $\omega_n^-$. It can be shown by direct calculation that the new
potential (\ref{v01}) also possess this set of solutions.
\vspace{.2cm}

{\underline{\bf Case 2. $V_0(x)$ with three NM's:}~ Let us now
consider the potential (\ref{qnm1}) supporting three NM's. A
convenient choice of the parameter is $\nu=-6.2$ so that $A^+=2.04,
A^-=-3.04$. We shall now construct the new potential using the NM
frequencies $\omega_1^+=1.04i$ and $\omega_2^+=0.04i$. The Wronskian
$W_{1,2}^+$ is found to be \beq W_{1,2}^+(x) =
\f{(sechx)^{2A^+-1}}{2(A^+-1)}[A^+-2-(A^+-1)cosh2x]\label{w03} \eeq
Now using (\ref{v2}) we obtain \beq V_2^+(x) =
-(A^+-1)(A^+-2)sech^2x + 8(A^+-1)\f{(A^+-2)cosh2x -
(A^+-1)}{[(A^+-1)cosh2x-(A^+-2)]^2}\label{v02} \eeq To get an idea
of the potential, we have plotted $V_2^+(x)$ in figure 2. From
figure 2, it is clear that $V_2^+(x)$ supports at least one NM. Next
to examine the wave functions we first consider $f^+(x)$ and
$g^+(x)$.
>From the relation (\ref{phiij}) we obtain \beq f^+(x)=
\f{2(A^+-1)(sechx)^{-A^+}tanhx}{(A^+-1)cosh2x-(A^+-2)}~~,~~g^+(x)=
\f{(sechx)^{-(A^++1)}[1-(2A^+-1)tanh^2x]}{(A^+-1)cosh2x-(A^+-2)}
\label{phi1212} \eeq Also from (\ref{phi}) it follows that \beq
\phi_0^+(x) =
\f{4(A^+-1)(sechx)^{(A^+-2)}}{(A^+-1)cosh2x+A^+-2}\label{phi120}
\eeq

>From (\ref{phi1212}) it follows that $f^+(x)$ and $g^+(x)$ are new QNM's
corresponding to frequencies $-\omega_2^+=-0.04i$ and
$-\omega_1^+=-1.04i$ respectively. The former has one node the later
has two nodes. The nodal structure of the QNM wave functions are
different from those obtained earlier. The reason for this is that
since we started with the first and second excited state NM's and
the Wronskian $W_{1,2}^+$ is nodeless, the behaviour of the original
wave functions $\psi_{1,2}^+(x)$ are retained by $f^+(x)$ and
$g^+(x)$. However, $\phi_0^+(x)$ is a NM at $\omega_0^+=2.04i$ and
it does not have a node because $\psi_0^+(x)$ does not have one.
Also other QNM wave functions $\phi_n^+(x), n=3,4,...$ have either
no node or one node. In figure 3 we have plotted the some of the
wave functions. We also note that although the potential in
(\ref{v01}) is of a similar nature as (\ref{qnm1}), the potential
(\ref{v02}) is of a completely different type. In particular it is a
non shape invariant potential. Finally we discuss the possibility of
a second set of solutions for the potential (\ref{v02}). We recall
that the existence of two sets of solutions for the potential
(\ref{qnm1}) (or (\ref{v01})) was due to the fact that the parameter
$\nu$ could be expressed as a product of two different parameters
$A^\pm$. However in the case of (\ref{v02}) the entire potential can
not be expressed in terms of two distinct parameters because of the
presence of the second term. Consequently the potential (\ref{v02})
has only one set of solution mentioned above.

\subsection{Construction of isospectral partner potential using QNM's}
{\bf \underline {Case 1. Potential based on consecutive QNM's:}}
Here we shall construct isospectral partner of a potential which has
only QNM's. Thus we consider $\nu=0.24$ and in this case
$A^\pm=-0.4,-0.6$. We consider the $A^+$ sector and begin with the
freequencies $\omega_0^+$ and $\omega_1^+$. In this case the
expression for the Wronskian $W_{0,1}^+$, the new potential
$V_2^+(x)$ and the QNM wave functions can be derived from the
expressions (\ref{w01}), (\ref{v01}) and (\ref{phin}) respectively
except that we now have to use a different parameter value. Thus the
new potential is given by \beq V_2^+(x) = -3.36~
sech^2x\label{v01q}\eeq For this potential the NM's corresponding to
$-\omega_0^+=0.4i$ and $-\omega_1^+=1.4i$ are given respectively by
\beq f^+= (sechx)^{1.4}~~,~~g^+= (sechx)^{0.4} tanhx\eeq Clearly
these NM's are not SUSY partner of any levels in $H_0$. The QNM's
are correspond to $\omega_n^+=-i(n+0.4), n=2,3,...$ and are given by
(\ref{phin}) with $A^+=-0.4$. We have plotted some of the wave
functions in fig 4. From the figure we find that the wave functions
$f^+(x)$ and $g^+(x)$ correspond to NM's and the other wave
functions represent QNM's which are the SUSY partners of the QNM's
in $H_0$. We note that as in (\ref{v01}) the potential (\ref{v01q})
has two sets of QNM's, the second of which corresponds to
$\omega_n^-$. \vspace{.2cm}

{\bf \underline {Case 2. Potential based on non consecutive QNM's:}}
Here we shall consider the previous parameter values (i.e, $A^+ =
-0.4$) and construct the new potential using the non consecutive
levels $\omega_0^+$ and $\omega_3^+$. In this case the Wronskian is
given by \beq W_{0,3}^+ =
\f{(sechx)^{(2A^+-3)}}{2(A^+-2)}[(9-6A^+)tanh^2x+3]\label{w03}\eeq
It can be shown that the Wronskian (\ref{w03}) is nodeless. Now
using the (\ref{v2}) the new potential is found to be \beq V_2^+(x) =
\f{(A^+-2)[2(A^+(A^+-2)(3A^+-7)-2A^+(A^+-1)(A^+-4)~cosh2x-(3-2A^+)^2(A^+-1)~sech^2x]}{[1-A^++(A^+-2)~cosh2x]^2}\label{v03}\eeq
The potential (\ref{v03}) is free of any singularity and is plotted
in fig 5. From figure 5 we find that it supports NM's. Also as
explained earlier, this potential has also one set of solution. We
now consider the wave functions corresponding to $\psi_0^+(x)$ and
$\psi_3^+(x)$. These are obtained from (\ref{phiij}) and are given
by \beq f^+(x) =
\f{2(A^+-2)}{(9-6A^+)tanh^2x+3}~(sechx)^{(3-A^+)}~~,~~g^+(x) =
\f{(1-2A^+)tanh^2x+3}{(9-6A^+)tanh^2x+3}~sinhx~(sechx)^{(1-A^+)}\label{f+g+}\eeq
The above wave functions (with zero and one node respectively)
represent NM's corresponding to $-\omega_3^+=3.4i$ and
$-\omega_0^+=0.4i$. The other wave functions can be obtained through
(\ref{phi}). The two QNM wave functions lying between $\omega_0^+$
and $\omega_3^+$ are $\phi_{1,2}^+(x)$ corresponding to
$\omega_1^+=-1.4i$ and $\omega_2^+=-2.4i$. We have plotted these
wave functions in fig 6. From figure 6, it can be seen that $f^+(x)$
and $g^+(x)$ are NM's while $\phi_{1,2}^+(x)$ are QNM's, with the
later having two nodes. The rest of the QNM wave functions
corresponding to the freequencies $\omega_n^+=-(n+0.4)i, n\neq 0,3$
are given by $\phi_n^+(x)$ and they have either zero or one node.

\section{Polynomial SUSY}
In first order SUSY, the anticommutator $\{Q,Q^\dagger\}$ of the
supercharges is a linear function of the Hamiltonian. On the other
hand in higher order SUSY, $\{Q,Q^\dagger\}$ is a nonlinear function
of the Hamiltonian. It will be shown here that the Hamiltonians
$H_0$ and $H_2$ are related by second order SUSY. To this end we
define the supercharges $Q$ and $Q^\dagger$ as follows: \beq Q =
\left(
\begin{array}{cc}
  0 & 0 \\
  L & 0  \\
\end{array}\rt)~~,~~Q^\dagger = \left(\begin{array}{cc}
  0 & L^\dagger \\
  0 & 0  \\
\end{array}\rt)\eeq where the operator $L$ is given by (\ref{L}).

Clearly the supercharges $Q$ and $Q^\dagger$ are nilpotent. We now
define a super Hamiltonian $H$ of the form \beq H = \left(
\begin{array}{cc}
  H_0 & 0 \\
  0 & H_2  \\
\end{array}\rt)
\eeq It can be easily verified that $Q, Q^\dagger$ and $H$ satisfy
the following relations : \beq [Q,H] = [Q^\dagger,H] = 0\eeq Then
the anticommutator of the supercharges $Q$ and $Q^\dagger$ is given
by a second order polynomial in $H$ : \beq H_{ss} = \{Q,Q^\dagger\}
= \left(\begin{array}{cc}
  L^\dagger L & 0 \\
  0 & LL^\dagger  \\
\end{array}\rt) = \left(H+\f{\delta}{2}\right)^2 - c{\cal I}\label{susy1}\eeq where ${\cal I}$ is the
$2\times 2$ unit matrix and \beq \delta = -(\omega_i^2 +
\omega_j^2)~~,~~c = \left(\f{\omega_i^2-\omega_j^2}{2}\right)^2 \eeq
Also we have \beq [Q,H_{ss}] = [Q^\dagger,H_{ss}] = 0
\label{susy2}\eeq The relations (\ref{susy1}) and (\ref{susy2})
constitute second order SUSY algebra.

As an example let us consider the potentials (\ref{qnm1}) and
(\ref{v01}). The corresponding Hamiltonians $H_0$ and $H_2$ are
obtained from (\ref{qnm}). In this case $\delta=0.5416$ and
$c=0.2916$ so that from (\ref{susy1}) we get \beq H_{ss} =  \left(H
+ 0.5416\right)^2 - 0.2916{\cal I}\eeq In a similar fashion one may
obtain $H_{ss}$ for the other pair of potentials.

\section{Conclusion} Here we applied the second order Darboux algorithm to the P\"oschl-Teller
potential and obtained new exactly solvable potentials admitting QNM
solutions. We have considered a number of possibilties to construct
the new potentials e.g, starting from NM's or starting from QNM's.
It has also been shown that the new potentials are related to the
original one by second order SUSY. We feel it would also be also
useful to analyse the construction of potentials using various
levels as well as for different values of the parameter $\nu$ (for
example, $\nu =$ half-integer) \cite{leung}. Finally we belive it
would be interesting to extend the present approach to other
effective potentials appearing in the study of Reissner-Nordstr\"om,
Kerr black hole etc.

\ed